# Direct Measurement of Piezoelectric Response around Ferroelectric Domain Walls in Crystals with Engineered Domain Configuration

Haiyan Guo, Alexei A. Bokov, and Zuo-Guang Ye

*Department of Chemistry, Simon Fraser University, Burnaby, British Columbia, V5A 1S6, Canada*

We report the first investigation of the piezoelectric response on a nanoscale in the poled ferroelectric crystals with engineered configuration of domains. Piezoresponse force microscopy of tetragonal $0.63Pb(Mg_{1/3}Nb_{2/3})O_3–0.37PbTiO_3$ relaxor-based ferroelectric crystals reviled that the $d_{33}$ piezoelectric coefficient is significantly reduced within the distance of ~1 μm from the uncharged engineered domain wall. This finding is essential for understanding the mechanisms of the giant piezoresponse in relaxor-based crystals and for designing new piezoelectric materials.

The discovery of giant piezoelectricity in relaxor ferroelectric single crystals,[1] which demonstrate the piezoelectric coefficients five to ten times as large as in traditional piezoelectric ceramics PZT, makes them the materials of choice for a new generation of electromechanical transducers, sensors and actuators [1,2,3,4] A great deal of effort has been made to investigate the origin of their special properties. The polarization rotation model [5,6] attributes the piezoresponse essentially to the intrinsic (crystal lattice strain) contribution. However, relaxor crystals possess complex polar structure on nano- and microscopic levels which may include local random fields, specific relaxor polar nanoregions, phase and composition heterogeneities, lamellar nanodomains of adaptive phase, ferroelectric microdomains, etc. [7] These features may in principle play an essential role in piezoelectric effect, giving rise to "extrinsic" piezoelectric contribution [8,9,10,11,12,13]. Experimentally, however, this role has not been clarified. As a result, understanding of the piezoelectricity mechanisms in relaxors is still lacking, which forms a serious obstacle for the development of new piezoelectric materials.

Significant macroscopic piezoresponse can be observed only in those ferroelectric samples which have been preliminarily poled by external electric field to get monodomain state or the state with the preferable domain orientation. In many cases (e. g. in ceramics) monodomain state cannot be attained and the sample consists of multiple domains separated by domain walls [14]. If the applied electric field or mechanical stress forms different angles with the spontaneous polarization vectors of adjacent domains, the wall between them can move and contribute thereby to piezoelectric effect. In conventional ceramic ferroelectric materials this contribution may be of the same order of magnitude as the intrinsic contribution [15]. However, besides the positive effect enhancing piezoelectric coefficients, domain wall motion leads to unwanted nonlinearity and hysteresis in the strain–field dependences. In the special case of so-called "engineered domain" configuration in crystals the walls are arranged in such way that the electric field or stress cannot influence their positions and the corresponding extrinsic piezoelectric contribution is absent. Surprisingly, the best piezoelectric performance is found right in crystals with engineered domain configuration including the relaxor-based perovskite solid solutions [1-4, 16]. To explain this fact it is tempting to assume that the regions near the walls possess enhanced piezoelectric properties even though the walls do not move [17,18]. In particular, it was predicted based on computer simulations that the possible mechanism of piezoresponse enhancment is the field-induced reversible phase transformation in the regions close to the wall [16]. Local measurement of piezoelectric coefficients needed for verifying this hypothesis have not been performed so far. These measurements cannot be done by conventional methods because of insufficient spatial resolution. In the present work to solve the problem we use the technique of piezoresponse force microscopy (PFM) which is expected to be a suitable tool because of its proved great capability in visualizing domain walls and measuring piezoelectric properties on the nanoscale [19]. We find that in the vicinity of engineered domain wall the piezoelectric response is significantly reduced, rather than enhanced as has been assumed until now.

Relaxor-ferroelectric (1-x)Pb(Mg$_{1/3}$Nb$_{2/3}$)O$_3$–$x$PbTiO$_3$ perovskite solid solution crystals grown by the Bridgman method were studied. The crystal of composition $x = 0.37$ (PMN-37PT) having tetragonal structure (crystal class 4$mm$) was chosen for investigation. At this composition the domain size is known to be much larger than in the rhombohedral and monoclinic crystals with smaller $x$ [20] (although piezoelectric coefficients are not so large) and consequently the piezoresponse from the internal parts of the domains and from the near-wall regions can be reliably separated.

Quantitatively the piezoelectric effect can be characterized with the matrix of piezoelectric coefficients, $d_{ij}$, through the relation

$$S_j = d_{ij}E_i, \qquad (1)$$

where $E_i$ is a component of the electric field vector and $S_j$ is a piezoelectric strain component. In ferroelectrics the piezoelectric coefficients are related to the spontaneous polarization vector, $\boldsymbol{P}_S$, through the expression

$$d_{ij} \approx \varepsilon_{im}Q_{jmk}P_{Sk}, \qquad (2)$$

where $\varepsilon_{im}$ and $Q_{jmk}$ are the dielectric constants and electrostriction coefficients, respectively.

To obtain engineered domain configuration in the crystal of tetragonal symmetry, the sample should be poled by a dc electric field applied along one of the <111> directions. Therefore, we prepared crystal plates with a thickness of about 150 µm, and the edges parallel to [111], [$\bar{2}$11] and [01$\bar{1}$] directions, respectively (it is a convention to use the crystallographic coordinate system of the parent cubic structure to characterize the ferroelectric domain states). The largest rectangular faces corresponding to the (111) planes were mirror-polished using a series of diamond pastes down to 1 µm and the gold electrodes were deposited on these faces by sputtering. To remove the stress caused by polishing, the samples were annealed at 500 °C for half an hour. The coercive field was estimated from ferroelectric hysteresis loops and appeared to be equal to 6 kV/cm. A much larger poling field of 10 kV/cm was used to ensure complete poling. The macroscopic piezoelectric coefficient $d_{33}$ measured by the conventional method using a Berlincourt $d_{33}$-meter at the frequency of 110 Hz appeared to be 460 pm/V, in agreement with the published data for the crystals of the same orientation and similar composition [21]

The $\boldsymbol{P}_S$ direction of any domain in a poled tetragonal crystal must form an angle of 90º with respect to the $\boldsymbol{P}_S$ of other domains; therefore the boundaries between domains are 90º-domain walls. The permissible directions of the walls can be calculated theoretically [22]. In crystals of 4mm symmetry obtained by cooling from the paraelectric phase of $m\bar{3}m$ symmetry (as in our case) the 90º domain walls must be parallel to one of the {110} crystallographic directions: all other walls would have larger elastic energy. The micrograph of a poled PMN-37PT crystal viewed along the <111> direction using polarizing microscope is shown in Fig. 1 (electrodes are removed). Figure 2 represents schematically the domain structure compatible with this micrograph and the theoretical predictions. Only the (101) walls are found. These walls must be uncharged (i.e. the $P_S$ vectors in adjacent domains are arranged in a head-to-tail manner) and inclined with respect to the crystal surface so that the domains have the form of wedges (charged walls would be arranged perpendicular to the $[01\bar{1}]$ direction). As a result, the wedge-shaped domains with different directions of optical indicatrix overlap each other on the way of the light and the parallel fringes of interference colors appear in the micrograph. The colour pattern should depend on the domain size and the number of domains through which the light propagates, which are unknown parameters. Therefore, the domain walls can hardly be distinguished. Nevertheless, the wall direction is evident: they are parallel to the colour fringes and form an angle of 60° with respect to the $[01\bar{1}]$ direction, as theoretically predicted.

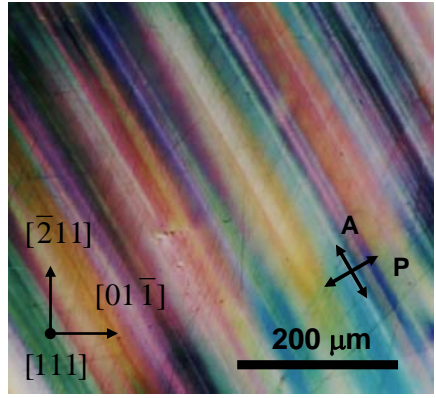

FIG. 1. Optical domain micrograph under polarizing microscope of the (111)-oriented poled PMN-37PT crystal plate. The crystallographic directions and the positions of crossed polarizer and analyzer are identified.

In contrast to optical microscopy, the PFM studies the domain structure in the regions close to the crystal surface. The PFM is a type of scanning probe microscopy [19] in which the piezoelectric strain is excited by the electric voltage applied between bottom electrode and the very sharp conducting PFM tip which touches the crystal surface (see Fig. 2). For technical reasons a high-frequency ac voltage must be used. Movement of the surface related to the piezoelectric strain leads to the displacement of the tip which can be measured. From these measurements, using equation (1) one can estimate the $d_{ij}$ values (in the laboratory, not the crystallographic, coordinate system) in a small region of the order of a few tens of nanometres. From equation (2) the components of $P_S$ can be estimated. The local $d_{33}$ coefficient determines the displacement of the tip in the vertical direction (index 33 means that piezoelectric deformation is in the same ($z$) direction as the applied electric field), while the $d_{15}$ coefficient (shear deformations of crystal) is mainly responsible for lateral displacement in the $x$ direction (perpendicular to cantilever) [19, 23]. By scanning the surface point by point the directions of $P_S$ in different areas (i.e. the domain configuration) can be mapped. Figure 3 presents the typical results of two-dimensional mapping of the same PNN-37PT crystal as shown in Fig. 1. The PFM experiments were performed on a modified atomic force microscope (Veeco ThermoMicroscopes). The PFM extension included a National Instrument (NI) Data Acquisition System (DAQ board), a lock-in amplifier and a LabView computer controlling system. An ac frequency of 95 kHz was used in the vertical PFM experiments. A 10 kHz ac frequency was used in the lateral PFM measurements in order to avoid possible tip slipping problem at high frequency [24]. The PFM images were developed with the help of WSxM software [25].

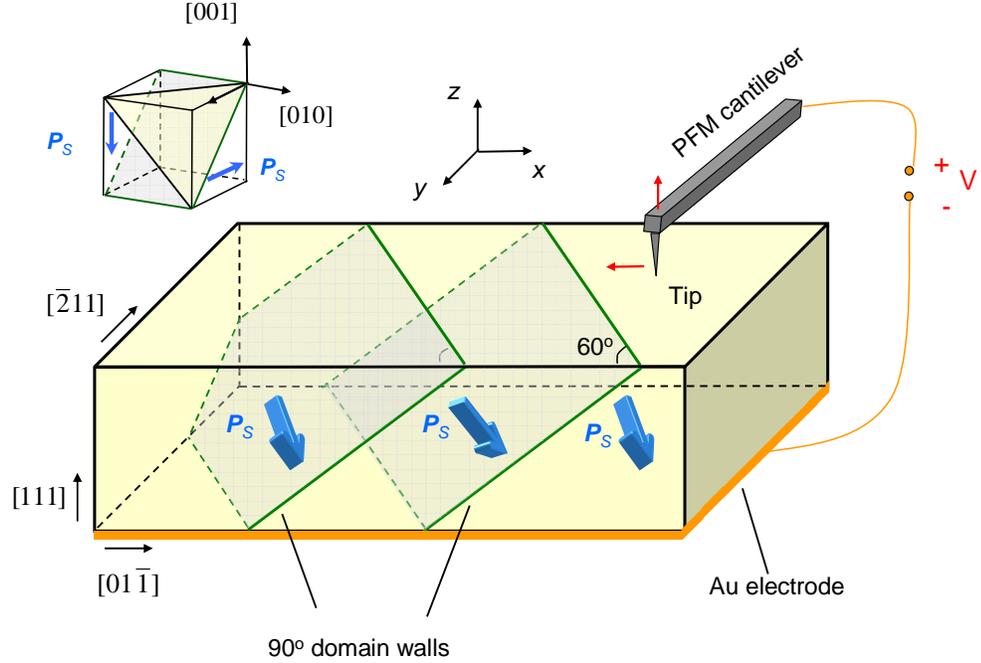

FIG. 2. Schematic diagram of the domain structure in the studied (111)-oriented poled PMN-37PT crystal plate and the experimental setup for PFM measurements (not to scale). Laboratory coordinate system (X, Y, Z), crystallographic directions and the positions of 90º domain walls are indicated. Blue arrows represent the spontaneous polarization vectors in different domains. Red arrows show the vertical and lateral displacements of the PFM tip when the measurement voltage (V) is applied. Inset shows the positions of the observed domain walls and crystal surface with respect to the cubic unit cell of the paraelectric phase.

The topography (which is obtained simultaneously with the PFM images at the same location) is shown in Fig. 3a (scratches from the polishing are visible). Fig. 3b shows the lateral (measured in the *X* direction) piezoresponse image. There are two large areas with comparatively small and large lateral PFM response, respectively. The relative integrated intensities in these two areas are shown in Fig. 4. In Fig. 3b the areas are separated by the straight boundary making an angle of 60º with the *X* direction (i.e. the [$01\bar{1}$] crystallographic direction). Similar boundaries are observed across the sample at a distance of 20-30 μm from one another. These boundaries are evidently the domain walls. Shear piezoelectric strain (and consequently lateral PFM response) is expected to be different in different domains because of different directions of the lateral components of the $P_S$ vector. In contrast, the vertical $P_S$ components are the same and, therefore, vertical PFM response should be the same in all domains. It is, indeed, practically the same both in phase (not shown here) and in amplitude (as shown in Fig. 3c). Nearly constant vertical response across the area is also clearly seen in Fig. 4, in which the integrated average value is shown as a function of distance from the wall. Note, however, that within ~ 1 μm distance of the domain wall the vertical response is significantly smaller than inside the domains. As discussed above, the vertical response is determined by the $d_{33}$ piezoelectric coefficient. Therefore, we conclude that in the vicinity of domain wall, the piezoelectric coefficient is significantly reduced, rather than enhanced as believed so far.

The origin of such a surprising behaviour of local $d_{33}$ across the domain walls can be related to the residual strains discovered around 90º walls in recent X-ray microdiffraction experiments [26]. The strain fields which were observed to extend several micrometers from the walls [27] may change piezoelectric coefficients considerably [11]. On the other hand, the charged walls, which are also permitted and can be observed in domain-engineered ferroelectric crystals, seem to enhance piezoelectric effect via the creation of depolarizing electric fields [27].

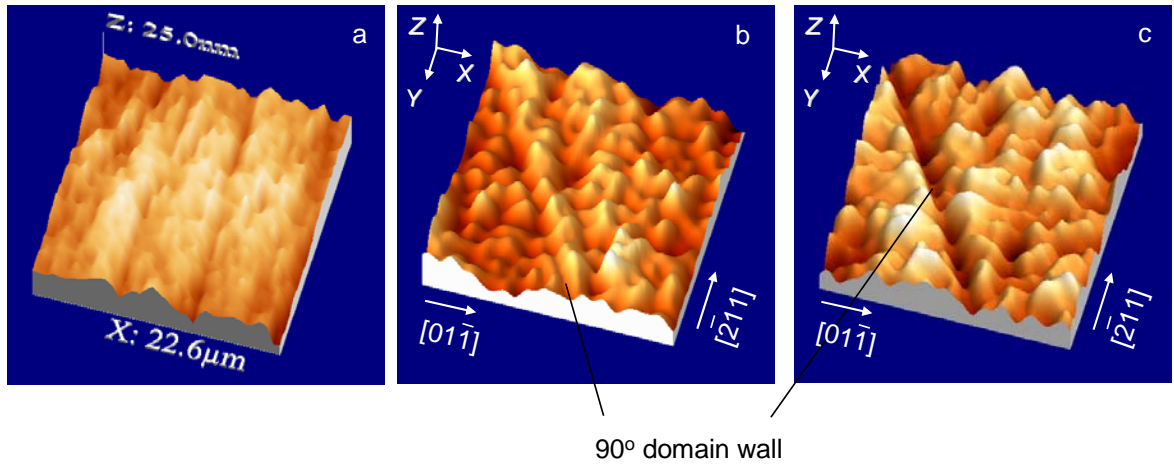

FIG. 3. PFM images obtained on a 23×23 μm surface area of the (111)-oriented poled PMN-37PT crystal around domain wall. a) Surface topography. b-c) Two-dimensional variations of lateral (b) and vertical (c) PFM response. *X* and *Y* axes represent the coordinates of the measurement point in the laboratory coordinate system; PFM response is plotted on the *Z* axis (in arbitrary units). The domain wall is identified.

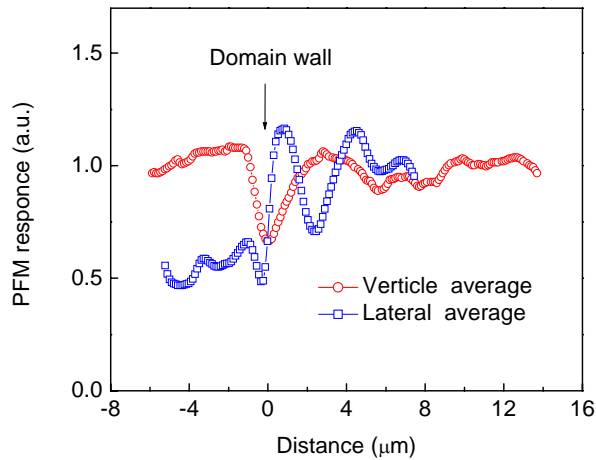

FIG. 4. PFM response in the vicinity of 90° domain wall in the (111)-oriented poled PMN-37PT crystal. Variation of vertical and lateral average values integrated in the direction parallel to the wall as a function of the distance to the wall. Scales for vertical and lateral response are different.

In conclusion, our results provide experimental evidence that uncharged domain walls have negative effects on the piezoelectric properties in the relaxor PMN-PT crystals with engineered domain structure as the $d_{33}$ piezoelectric coefficient significantly decreases in the vicinity of the walls. To increase macroscopic $d_{33}$ it would therefore be necessary to fabricate piezoelectric crystals oriented and poled along a nonpolar direction but free from uncharged walls; the method of such fabrication process needs to be developed.

The authors acknowledge S. V. Kalinin and V. Yu. Topolov for helpful discussions and H. Luo for preparing the PMN-PT crystals. This work was supported by the U.S. Office of Naval Research (Grant No. N00014-06-1-0166) and the Natural Science and Engineering Research Council of Canada (NSERC).